\documentclass[11pt,preprint]{aastex}

\def\hh{\hbox{H$_2$}}

\slugcomment{}

\begin{document}

\title{Diffuse Far-ultraviolet Observations of the Taurus Region}

\author{
D. H. Lee\altaffilmark{1}, I. S. Yuk\altaffilmark{1}, H.
Jin\altaffilmark{1}, K. I. Seon\altaffilmark{1}, J.
Edelstein\altaffilmark{2}, E. J. Korpela\altaffilmark{2}, J.
Adolfo\altaffilmark{2}, K. W. Min\altaffilmark{3}, K. S.
Ryu\altaffilmark{3}, J. H. Shinn\altaffilmark{3},  and E. F. van
Dishoeck\altaffilmark{4} }

\email{email: dhlee@kasi.re.kr}

\altaffiltext{1}{Korea Astronomy and Space Science Institute,
Daejeon, Korea, 305-348; email: dhlee@kasi.re.kr, yukis@kasi.re.kr,
jinho@kasi.re.kr, kiseon@kasi.re.kr} \altaffiltext{2}{Space Sciences
Laboratory, University of California, Berkeley, CA, USA 94720;
email: jerrye@ssl.berkeley.edu, korpela@ssl.berkeley.edu,
adolfo@ssl.berkeley.edu} \altaffiltext{3}{Korea Advanced Institute
of Science and Technology, Daejeon, Korea 305-701; email:
kwmin@space.kaist.ac.kr, ksryu@space.kaist.ac.kr,
jhshinn@space.kaist.ac.kr} \altaffiltext{4}{Sterrewacht Leiden,
Postbus 9513, 2300 RA Leiden, The Netherlands; email:
ewine@strw.leidenuniv.nl}

\begin{abstract}

Diffuse far-ultraviolet (FUV: 1370--1670 \AA) flux from the Taurus
molecular cloud region has been observed with the \emph{SPEAR/FIMS}
imaging spectrograph. An FUV continuum map of the Taurus region,
similar to the visual extinction maps, shows a distinct cloud core
and halo region. The dense cloud core, where the visual extinction
is $A_v$ $>$ 1.5, obscures the background diffuse FUV radiation,
while a scattered FUV radiation is seen in and beyond the halo
region where $A_v$ $<$ 1.5. The total intensity of \hh~fluorescence
in the cloud halo is $I_{\hh} = 6.5$ $\times 10^4$ photons cm$^{-2}$
s$^{-1}$ sr$^{-1}$ in the 1370--1670 \AA~wavelength band. A
synthetic model of the \hh~fluorescent emission fits the present
observation best with a hydrogen density $n_{\rm H}$ = 50 cm$^{-3}$,
\hh~column density $N$(\hh) = 0.8 $\times 10^{20}$ cm$^{-2}$, and an
incident FUV intensity $I_{\rm UV}$ = 0.2. \hh~fluorescence is not
seen in the core presumably because the required radiation flux to
induce fluorescence is unable to penetrate the core region.

\end{abstract}


\section{INTRODUCTION}

There are two major components in the Galactic diffuse
far-ultraviolet (FUV) emission: scattered starlight from
interstellar dust and \hh~fluorescent emissions (Bowyer 1991). Both
the diffuse FUV and infrared (IR) backgrounds are important tracers
of interstellar dust. Much of the current information on the
large-scale distribution of dust has been obtained from IR missions
such as {\it IRAS} and {\it COBE} (see Sodroski et al. 1997). Luhman
et al. (1994) observed vibrationally excited \hh~emissions in the
near-IR, and Luhman \& Jaffe (1996, hereafter LJ96) argued that
there was a significant correlation between the \hh~line intensity
and the far-IR (FIR) continuum, the origin of which they attributed
to the ultraviolet illumination of molecular hydrogens and dust on
cloud surfaces. Scattering in the FUV wavelengths may also give
complementary information to that of the IR emission, and the
combination of these two wavebands can lead to a unique
determination of interstellar dust parameters (Murthy \& Sahnow
2004, hereafter MS04). Haikala et al. (1995) observed a Galactic
cirrus cloud near the north Galactic pole with the FUV imaging
telescope FAUST, and showed a good correlation between the FUV and
{\it IRAS} 100 $\mu$m surface brightness. Based on the {\it FUSE}
observations, MS04 argued that even at low Galactic latitudes, the
diffuse FUV ($<$ 1200 \AA) sky should be patchy with regions of
intense continuum, mostly near bright stars, and dark regions. They
found a weak correlation between the FUV flux and the 100 $\mu$m
emissions, but with large variations. Hurwitz, Bowyer \& Martin
(1991) pointed out that for a fully clumped dust model, the
scattered FUV continuum should fall as the FUV optical depth exceeds
$\sim$ 1, while the 100 $\mu$m radiation continues to increase with
a neutral hydrogen column density.

FUV photons play an important role in the physical and chemical
processes in the interstellar medium (ISM). They ionize atoms,
dissociate molecules, and heat gases both by ejecting electrons from
dust grains and by directly exciting atoms and molecules (see, e.g.,
Tielens \& Hollenbach 1985; Black \& van Dishoeck 1987, hereafter
BvD87; van Dishoeck \& Black 1988; Sternberg 1989). Hydrogen
molecules are electronically excited by absorbing FUV photons in the
Lyman and Werner bands while their spontaneous decay back to the
electronic ground states results in the emission of FUV lines,
cascading down to the lower vibration-rotation levels by emitting
near-IR lines (BvD87; Luhman et al. 1997). Since the FUV radiation
field in a cloud decreases rapidly with an increasing optical depth,
both the abundance and the excitation of \hh~are significantly
optical-depth-dependent (BvD87), which enables us to model the
physical conditions of a cloud from the observation of the \hh~UV
fluorescence (Witt et al. 1989). Molecular hydrogen fluorescent
emissions in the IR bands have been observed from various
star-forming regions, such as the Galactic center (Pak, Jaffe \&
Keller 1996) and the Orion nebula (Luhman et al. 1994). In the FUV
bands, \hh~fluorescence was first discovered near IC 63, which
resides in the intense FUV radiation field of a nearby B-star (Witt
et al. 1989). \hh~fluorescence in the diffuse interstellar medium
excited by a general interstellar radiation field was first detected
by Martin, Hurwitz \& Bowyer (1990).

The Taurus region is a local complex of molecular clouds where
active star formation is in progress. With a relatively short
distance of 140 pc from Earth (Kenyon et al. 1994) and being
sufficiently far from the Galactic plane, it suffers negligible
foreground and background extinction (Padoan, Cambresy \& Langer
2002, hereafter PCL02). Hence, the Taurus cloud region is a useful
laboratory for understanding the relationship between interstellar
gas, dust, and FUV radiation. The FUV continuum and \hh~fluorescence
spectra ($\Delta\lambda \sim 10$ \AA) have been obtained from a
0.13$^{\circ}$ $\times$ 3.8$^{\circ}$ field in the Taurus region and
analyzed in detail (Hurwitz 1994).

In this paper, we report the results of new observations in the
Taurus region taken with the FUV imaging spectrograph, {\it
SPEAR/FIMS}, onboard the Korean micro-satellite STSAT-1, launched on
September 27, 2003. Particularly, we present the first FUV continuum
image (10$^{\circ}$ $\times$ 24$^{\circ}$) of the Taurus region
mapping the scattered starlight from the dust. The FUV continuum
image is compared with the visual extinction map, as well as with
the FIR emission map. The observations also reveal that H$_2$
fluorescent emission is present in the halo regions, but not in the
core region. The physical parameters of the H$_2$-fluorescent region
are obtained using the synthetic \hh~model developed by BvD87.

\section{FAR-ULTRAVIOLET OBSERVATIONS AND ANALYSIS}

The {\it SPEAR/FIMS} mission employs an imaging spectrograph with
two channels (Short and Long) which is optimized for the observation
of diffuse emission lines at FUV wavelengths. We utilized the Long
channel data only in this study, while the analysis of the Short
channel ($\lambda\lambda$ 900--1150 \AA) data is proceeding. The
Long channel covers $\lambda\lambda$ 1350--1700 \AA, with a
resolving power of $\lambda/\Delta\lambda \sim$ 550. The field of
view of the Long channel is $7.4^\circ \times 4.3'$, with a 5--10$'$
imaging resolution along the slit, and the spectral half-energy line
width, averaged over the angular field, is 3.2 \AA. The instrument,
its in-orbit performance, and the data analysis procedures are
described in Edelstein et al. (2005a, b).

Our data set is composed of 21 scanning observations made during the
sky survey and 3 orbits of a dedicated pointing observation toward
the Taurus cloud. The total exposure time of the 21 scanning orbits
is around 300 seconds per pixel, on average, while that of the
pointing observations is 1629 seconds. The internal detector
background, measured during the observation for 5 s at 25 s
intervals by closing the shutter, is 0.01 counts s$^{-1}$ \AA$^{-1}$
and subtracted from the data. Only the data from 1370 to 1670 \AA,
excluding the intense O {\scriptsize I} airglow line at 1356 \AA,
are used for the present analysis.

Figure 1 shows the observed FUV (1370--1670 \AA) continuum map with
0.2$^\circ$ $\times$ 0.2$^\circ$ pixels, smoothed with 3 pixels to
reduce statistical fluctuations. The two narrow rectangles represent
the fields of the present pointed observation and the UVX target of
Hurwitz (1994) centered at ($l,b$) = (168, -16). Bright {\it TD}-1
catalog stars ($> 2 \times 10^{12}$ ergs cm$^{-2}$ s$^{-1}$
sr$^{-1}$ \AA$^{-1}$) from Gondhalekar, Philips \& Wilson (1980) are
also marked. It is apparent that the map is divided into distinctive
FUV intensity regions, represented by colors: the blue region with
0--1250 photons cm$^{-2}$ s$^{-1}$ sr$^{-1}$ (hereafter CU)
corresponds to the core, the red (1250--2500 CU) to the halo region,
and the white ($>$ 2500 CU) to the stars and the diffuse background
region beyond the Taurus cloud. We have also overplotted the
contours of the visual extinction obtained from the map in PCL02:
(1) $1.6 < A_v < 19.6$, (2) $0.4 < A_v < 1.6$, and (3) $A_v < 0.4$.
We can easily note that the FUV intensity is well related with the
contours of $A_v$. The FUV intensity of each pixel may have an
uncertainty of 25 \% due to the systematic error in estimating the
effective area (see Edelstein et al. [2005a] for instrumental
issues). Nevertheless, we believe this continuum map certainly shows
that the Taurus cloud is obscuring a more distant diffuse FUV
background source, consistent with the picture of Hurwitz (1994),
whose results were based on the limited UVX observation. We will
discuss this further in the next section.

The pointed observation includes the Taurus halo region (field {\it
a}) as well as the core region (field {\it b}). Fields {\it a} and
{\it b} are 4.3$' \times$1$^\circ$ and 4.3$' \times$ 1.7$^\circ$,
and centered at ($l,b$) = (167.3, -17.3) and (168.1, -16.2),
respectively. Figure 2 shows the spectra obtained from these two
regions. Each spectrum is made with 1 \AA~bins and a boxcar smoothed
by 3 bins. The Poisson statistical uncertainty levels are also
indicated. It is clear from this figure that the \hh~fluorescence is
significant in the halo region (field {\it a}), while it is not
significant in the core region (field {\it b}). It should also be
noted that the systematic uncertainty in the effective area may
affect the overall intensity but not the spectral shape.

We modeled the \hh~fluorescence spectrum of the halo region using
CLOUD, a plane parallel \hh~model program for the photo-dissociation
regions (PDRs). The basic concepts, physical backgrounds, and the
application of the model are described in BvD87. While many input
parameters were required for detailed modeling, we focus on only
three main parameters: the enhancement factor of the incident FUV
intensity ($I_{\rm UV}$) compared with the mean interstellar value
adopted by Draine (1978), the cloud density $n_{\rm H}$, and the
total \hh~column density $N$(\hh). We have generated 3200 synthetic
models with $n_{\rm H}$ = (10, 50, 100, 500) cm$^{-3}$, $N$(\hh) =
(0.1 $\sim$ 20) $\times$ 10$^{20}$ cm$^{-2}$, and $I_{\rm UV}$ =
(0.01 $\sim$ 2) to find the best fit to the observed FUV
fluorescence. We use the spectrum of the core region (smoothed by 21
bins and scaled) as a model continuum assuming that the spectral
shapes of the background continuum in the core region and in the
halo region are similar. Though this might be a source of
uncertainty, it should be noted that no significant spectral
features are apparent in the spectrum of the core region (field {\it
b}). The scale factor of the model continuum and the cloud
parameters were chosen to minimize the $\chi^2$ value. The final
result of the model continuum (solid line) and the fit (dashed line)
is overplotted in Figure 2a. The parameters we obtained for the halo
region are $n_{\rm H}$ = 50 cm$^{-3}$, $N$(\hh) = 0.8 $\times$
10$^{20}$ cm$^{-2}$, and $I_{\rm UV}$ = 0.2, with $\chi^2_\nu$ =
0.899. The corresponding molecular hydrogen fluorescent intensity in
the 1370--1670 \AA~wavelength band is 6.5 $\times 10^4$ photons
cm$^{-2}$ s$^{-1}$ sr$^{-1}$ or 8.4 $\times 10^{-7}$ ergs cm$^{-2}$
s$^{-1}$ sr$^{-1}$. The current results are generally consistent
with those obtained by Neufeld \& Spaans (1996, hereafter NS96) for
the UVX 1400--1700 \AA~data: $n_{\rm H}$ = 50 cm$^{-3}$ and $I_{\rm
UV}$ = 0.4 for the observed fluorescence intensity of 7 $\times
10^{-7}$ ergs cm$^{-2}$s$^{-1}$ sr$^{-1}$. Although $I_{\rm UV}$ is
different by a factor of 2, the increase of the enhancement factor
$I_{\rm UV}$ from 0.2 to 0.4 changes $\chi^2$ by $<$ 10 \% in our
model because the absolute FUV flux increment is small and the
increased $I_{\rm UV}$ is offset by the decrease of the fit
background. When we apply a flat background for the model continuum
instead of the core spectrum, the best fit results in similar
parameters within 10 \% variations, except $I_{\rm UV}$, which
increases 2.5 times, with the $\chi^2_\nu$ value 0.713.

\section{DISCUSSION}

Observations in different wavebands of the Taurus region show
similar pictures that the clouds appear as diffuse filaments with
dense clumps of molecular gas embedded (Whittet et al. 2004, and
references therein). The FUV map shown in Figure 1, together with
the 2MASS visual extinction map in PCL02, and the FIR maps of
$I_{100{\mu}{\rm m}}$ and $I_{60{\mu}{\rm m}}$ (see Abergel et al.
1994), provide an excellent opportunity for a comparative study of
the dust scattering observed in different wavelength bands. The FUV
color map clearly shows an anti-correlated structure when compared
with the visual extinction map and the cold component map
($I_{100{\mu}{\rm m}} - I_{60{\mu}{\rm m}}/0.15$) of Abergel et al.
(1994): the FUV intensity decreases toward the Taurus core region
while the visual extinction and the FIR intensity increase. This
anti-correlation is contradictory to the general correlation between
FUV and IR in optically thin ($A_v < 1$) regions (Hurwitz 1994;
LJ96).

To further explore the relationship between the FUV continuum
intensity $I_{\rm FUV}$ and the visual extinction $A_{v}$, we
rescaled the visual extinction map to 0.2$^{\circ}$ $\times$
0.2$^{\circ}$ pixels to match our FUV map and compared the maps
pixel by pixel in Figure 3. The visual extinction used has an
uncertainty of 0.49 magnitude, according to PCL02, which causes $A_v
< 0$ for some data points, as seen in the figure. As expected,
$I_{\rm FUV}$ is low for high extinctions ($A_{v} >$ 1.5) since the
clouds block the background radiation. Also, $I_{\rm FUV}$ is more
or less flat in this high extinction region, which indicates $I_{\rm
FUV}$ comes mainly from the foreground radiation and its scattered
light. For low extinctions, $A_{v} <$ 1.5, on the other hand, the
present result shows scattered $I_{\rm FUV}$. This is inconsistent
with the general notion that $I_{\rm FUV}$ increases with $A_{v}$
due to dust scattering in optically thin conditions. We believe the
ambiguity in the present case arises as the scattered photons of the
foreground FUV lights are mixed with those of the background; though
the uncertainty of $A_{v}$ might also have some effects.
Nevertheless, it should be noted that some of the data points used
by Murthy \& Sahnow (2004) to study the relationship between $I_{\rm
FUV}$ and $I_{100{\mu}{\rm m}}$ deviate significantly from the
simple linear relationship, which may be more appropriate for
optically thin regions. This is not unexpected as the study was
based on the $\it {FUSE}$ measurements that include targets in the
region where $I_{100{\mu}{\rm m}}$ exceeds $\sim$ 10 MJy, and
therefore, $A_v$ $>$ 1 (Hurwitz, Bowyer \& Martin 1991). Figure 3
shows that the relation between $I_{\rm FUV}$ and $A_{v}$ should be
compound when moving from an optically thin region to an optically
thick region. The current observation of FUV intensity over a broad
range of opacities within a single field, where the local
interstellar radiation field should be fairly constant, provides a
valuable data set that we intend to explore further in future
studies by using more detailed 3-dimensional optical-transfer
modeling to study the scattering properties of dust grain, as well
as the nature of the foreground and background illumination.

The present observation shows a more intense FUV continuum averaged
over 1370--1670 \AA~in the halo region (1079 $\pm$ 375 CU) than in
the core region (769 $\pm$ 267 CU), as Hurwitz (1994) described it:
a dark core with a bright rim structure (based on limited UVX
observations). The contours in Figure 1 indicate that the visual
extinction of the core region is about 15 times larger than that of
the halo region. Assuming a simple linear relationship between the
visual extinction and the hydrogen column density, this implies that
there is about 15 times more gas in the core region than in the halo
region. If we further assume that the dust FUV scattering properties
are constant over the Taurus region, we expect the incident FUV
intensity in the core region to be 21 times lower than that of the
halo region, with $I_{\rm UV} \sim$ 0.01, which is insufficient to
excite significant \hh~fluorescence.

\section{CONCLUSION}

We observed the FUV (1370--1670 \AA) emission from the Taurus region
using the FUV imaging spectrograph, {\it SPEAR/FIMS}, onboard the
Korean micro-satellite STSAT-1. Our map of the FUV continuum is
consistent with the picture in which the cloud both obscures a more
distant diffuse background source, as well as scatters the
foreground radiation. We found \hh~fluorescence only from the
cloud's halo, not from its core region, as the incident FUV
intensity in the core region is not sufficient to excite significant
\hh~fluorescence. A simple plane parallel \hh~model fits the halo
spectrum best with $I_{\rm \hh}$ = 6.5 $\times 10^4$ photons
cm$^{-2}$ s$^{-1}$ sr$^{-1}$, the cloud density $n_{\rm H}$ = 50
cm$^{-3}$, \hh~column density $N$(\hh) = 0.8 $\times$ 10$^{20}$
cm$^{-2}$, and the incident FUV intensity $I_{\rm UV}$ = 0.2. It is
seen that a mixed relationship exists between $I_{\rm FUV}$ and
$A_{v}$ with a transition point at ($A_{v} \sim 1.5$), dividing
optically thin regions from optically thick regions in the FUV band.

We thank Dr. Paolo Padoan and Dr. Laurent Cambresy who provided the
raw data of the Taurus regions visual extinction map for this
analysis. This publication makes use of data products from {\it
SPEAR/FIMS}, which is a joint project of Korea Astronomy and Space
Science Institute, Korea Advanced Institute of Science and
Technology, and University of California at Berkeley, funded by the
Ministry of Science and Technology (Korea) and the National
Aeronautics and Space Administration (USA).

\clearpage

\begin{figure}
\begin{center}
\includegraphics[scale=.5]{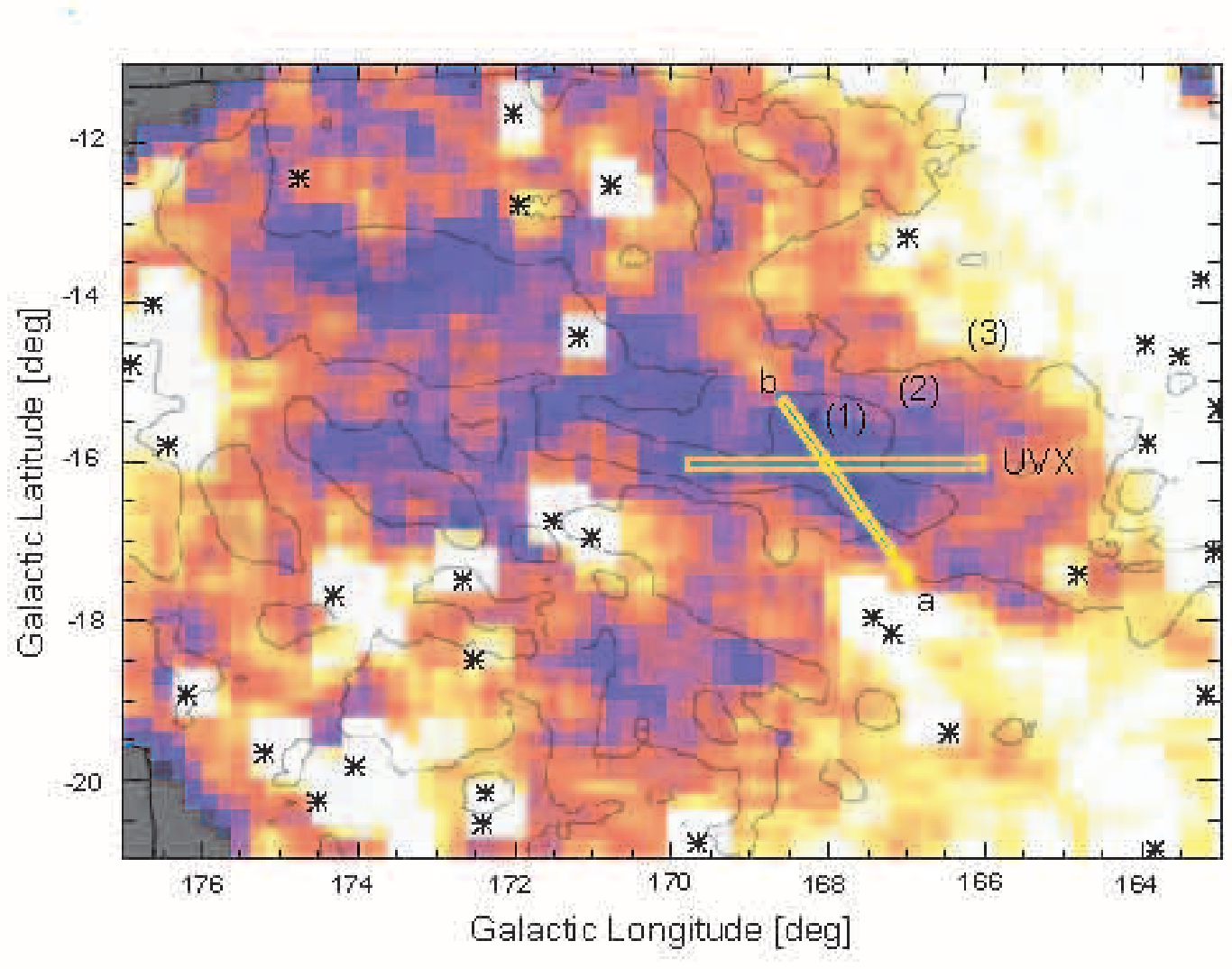}
\end{center}
\caption{FUV (1370--1670 \AA) map of the Taurus region obtained from
the survey observations by {\it SPEAR/FIMS}. The pixel size is
0.2$^\circ \times 0.2^\circ$ and the map is smoothed by 3 pixels to
reduce statistical fluctuations. The colors represent the FUV
intensity: blue (0--1250 CU), red (1250--2500 CU), and white ($>$
2500 CU), while the contours represent the distinct $A_v$ regions:
(1) $1.6 < A_v < 19.6$, (2) $0.4 < A_v < 1.6$, and (3) $A_v < 0.4$.
The bright {\it TD}-1 catalog stars ($ > 2 \times 10^{12}$ ergs
cm$^{-2}$ s$^{-1}$ sr$^{-1}$ \AA$^{-1}$) are overplotted on the
figure. The UVX and {\it SPEAR/FIMS} pointed observation fields
({\it a} and {\it b}) are also shown by the yellow rectangles.}
\label{fig1}
\end{figure}

\clearpage

\begin{figure}
\begin{center}
\includegraphics[scale=.5]{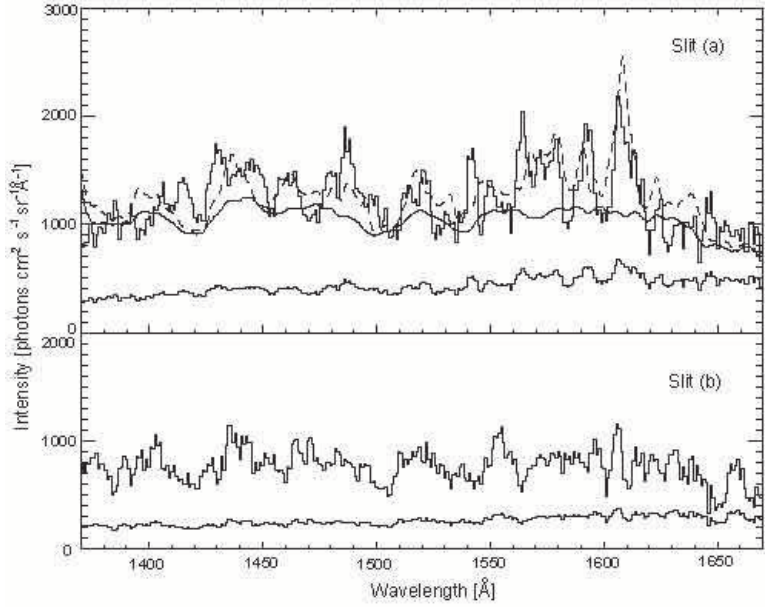}
\end{center}
\caption{FUV spectra (solid histograms) of the fields {\it a} (upper
panel) and {\it b} (lower panel), corresponding to the halo and core
regions, respectively. Each spectrum is obtained with 1 \AA~bin and
smoothed by 3 bins for optimal display. The model continuum and the
fit with \hh~fluorescence for the halo region are overplotted as
solid and dashed lines, respectively. The calculated uncertainties
(statistical errors: see text) are also shown near the bottom of
each panel.} \label{fig2}
\end{figure}

\clearpage

\begin{figure}
\begin{center}
\includegraphics[scale=.5]{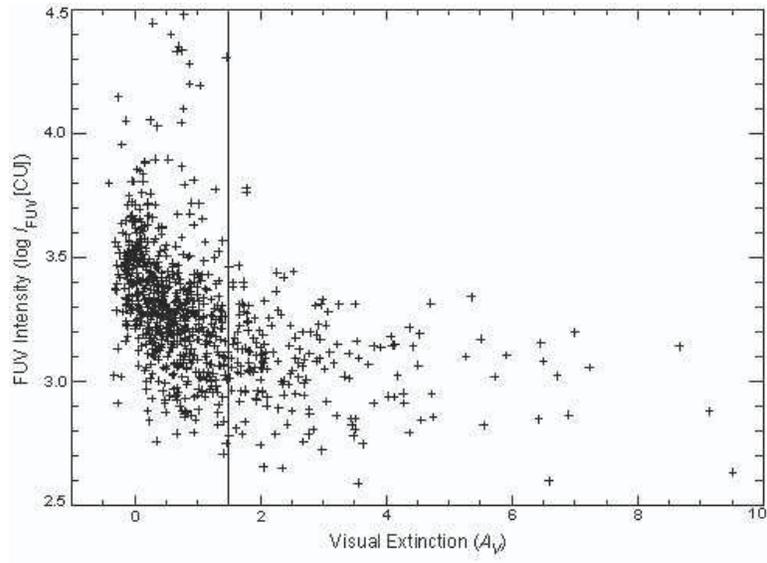}
\end{center}
\caption{A pixel-to-pixel diagram obtained from the visual
extinction map and the FUV continuum intensity map. The visual
extinction map was resized by 0.2$^{\circ}$ $\times$ 0.2$^{\circ}$
pixels for comparison.} \label{fig3}
\end{figure}

\end{document}